\documentclass[a4paper,twocolumn,english,aps,prl,superscriptaddress]{revtex4-1}
\usepackage[T1]{fontenc}
\usepackage[latin9]{inputenc}
\usepackage{color}
\usepackage{babel}
\usepackage{amsmath}
\usepackage{amssymb}
\usepackage{graphicx}
\DeclareGraphicsExtensions{.pdf}
\usepackage{amsmath,amssymb,bbold,bm,color}
\usepackage{float}
\usepackage{epstopdf}
\usepackage{tabularx}
\usepackage{braket}
\usepackage{booktabs}
\usepackage{array}
\usepackage{blindtext}
\usepackage{natbib}
\usepackage[unicode=true,pdfusetitle,
 bookmarks=true,bookmarksnumbered=false,bookmarksopen=false,
 breaklinks=false,pdfborder={0 0 1},backref=false,colorlinks=false]
 {hyperref}

\usepackage{ulem} 

\makeatletter

\usepackage{babel}
\usepackage{babel}

\begin{document}

\title{Non-Bloch Quantum Geometry of Non-Hermitian Systems}

\author{Junsong Sun}
\affiliation{Department of Physics and Astronomy, Seoul National University, Seoul 08826, Korea}
\affiliation{School of Physics, Beihang University,
Beijing, 100191, China}

\author{Huaiming Guo}
\email{hmguo@buaa.edu.cn}
\affiliation{School of Physics, Beihang University,
Beijing, 100191, China}

\author{Bohm-Jung Yang}
\email{bjyang@snu.ac.kr}
\affiliation{Department of Physics and Astronomy, Seoul National University, Seoul 08826, Korea}
\affiliation{Center for Theoretical Physics (CTP), Seoul National University, Seoul 08826, Korea}
\affiliation{Institute of Applied Physics, Seoul National University, Seoul 08826, Korea}

\begin{abstract}
We formulate quantum geometry for non-Hermitian systems under open boundary conditions. By defining quantum-geometric quantities in both real-space and non-Bloch representations, we establish a unified framework beyond conventional Bloch band theory. Our central result is an exact equivalence between the real-space integrated quantum metric and a non-Bloch integrated quantum metric defined on the generalized Brillouin zone. We further introduce localized non-Bloch Wannier functions in the presence of the non-Hermitian skin effect and show that the non-Bloch integrated quantum metric gives the gauge-invariant part of their spread functional. These results establish quantum geometry as a natural framework for characterizing open-boundary non-Hermitian band structures and the localization properties encoded in skin modes.
\end{abstract}

\date{\today}

\maketitle

Quantum geometry has emerged as a fundamental framework for characterizing the structure of Bloch states in crystalline systems~\cite{Resta2011, Torma2022, PhysRevLett.131.240001, liutianyu2024,yu2025quantumgeometryquantummaterials}. Within Bloch band theory, geometric quantities such as the Berry curvature and quantum metric play central roles in a wide range of physical phenomena, including polarization, Landau level spectra, nonlinear transport, superfluid weight, and topological responses~\cite{gao2023quantum,wang2023quantum,tian2023evidence,rhim2020quantum,hwang2021geometric,PhysRevB.109.035134,kang2025measurements,doi:10.1126/science.ado6049, sun2025GeoSFQC}. This geometric framework, however, presupposes the validity of Bloch band theory. In non-Hermitian systems under open boundary conditions, Bloch band theory generally breaks down~\cite{PhysRevLett.116.133903, Xiong_2018} because of the non-Hermitian skin effect~\cite{PhysRevLett.77.570,Li2020,Zhang31122022, Zhu_2024, PhysRevB.106.075112, Lin2023, PhysRevB.97.045106}, in which bulk eigenstates are no longer described by extended Bloch waves but instead become non-Bloch waves localized near the boundary~\cite{PhysRevLett.121.086803, ZhangXintong_2024, PhysRevB.100.035102, ZhangKai2025}. Consequently, the conventional bulk-boundary correspondence must be reformulated in terms of non-Bloch band theory defined on the generalized Brillouin zone (GBZ)~\cite{PhysRevLett.121.086803, PhysRevLett.123.246801, PhysRevLett.123.066404, PhysRevLett.125.226402, PhysRevLett.121.136802, PhysRevX.14.021011}. Although substantial progress has been made in understanding non-Hermitian spectral topology and bulk-boundary correspondence within this framework~\cite{wang2025, PhysRevLett.132.050402, Kaneshiro2025, PhysRevLett.121.136802}, the corresponding theory of quantum geometry under open boundary conditions remains largely undeveloped.

Existing studies of non-Hermitian quantum geometry have mainly considered periodic boundary conditions, where Bloch-like descriptions remain applicable~\cite{QMnHSSH_PRR,PhysRevResearch.7.L012067,behrends2025NHQM, QM_Ozawa, NHQG_Milosz}. Such approaches, however, do not capture the intrinsically non-Bloch nature of open-boundary eigenstates, which lies at the heart of non-Hermitian physics. Despite recent progress~\cite{NHQG_Kawabata}, the physical meaning of non-Hermitian quantum geometry remains unclear, especially its relation to the spatial structure of eigenstates and to non-Bloch band theory.

In this work, we address this problem by developing a general framework for quantum geometry in non-Hermitian systems under open boundary conditions. Our central results are twofold. First, we show that the real-space integrated quantum metric (IQM) under open-boundary conditions is exactly equivalent to a non-Bloch IQM defined on the GBZ. Second, we construct localized non-Bloch Wannier functions and show that this IQM gives the gauge-invariant part of their spread functional. Thus, the non-Bloch IQM is not merely a formal extension of the Bloch quantum metric, but directly characterizes the localization properties of non-Bloch Wannier functions. Our results provide a quantum-geometric foundation for open-boundary non-Hermitian systems and reveal how the spatial structure of skin modes is encoded in non-Bloch quantum geometry.

\textit{Real-space quantum geometry under open-boundaries.---}
We begin by formulating quantum geometry directly in real space. This approach is natural for non-Hermitian systems with the skin effect, where both the energy spectrum and the spatial structure of eigenstates under open-boundary conditions can differ drastically from their periodic-boundary counterparts. For clarity, we focus on one-dimensional (1D) systems.

We introduce the following non-Hermitian generalization of the real-space IQM:
\begin{align}\label{eqRIQGT}
	\mathcal{Q}^{\rm rs}_{m,xx}
	=
	-\frac{1}{N}{\rm Tr}
	\left\{
	\hat{P}_m
	\left[\hat{x},\hat{P}_m\right]
	\left[\hat{x},\hat{P}_m\right]
	\right\},
\end{align}
where \(N\) is the system size, \(\hat{x}\) is the position operator, and
\(\hat{P}_m=\sum_i|\psi^R_{m,i}\rangle\langle\psi^L_{m,i}|\)
is the projector onto the target energy sector labeled by \(m\) of the real-space non-Hermitian Hamiltonian \(H\). Here, \(i\) indexes distinct bulk eigenstates within the \(m\) sector, excluding topological boundary states. The right and left eigenvectors satisfy
\(H|\psi_{m,i}^R\rangle = E_{m,i}|\psi_{m,i}^R\rangle\) and
\(\langle\psi^L_{m,i}|H = E_{m,i}\langle\psi^L_{m,i}|\),
together with the biorthonormality condition
\(\langle \psi_{m,i}^L | \psi_{n,j}^R \rangle = \delta_{mn}\delta_{ij}\).
Equation~(\ref{eqRIQGT}) is analogous to the real-space IQM in Hermitian systems~\cite{PhysRevLett.122.166602, PhysRevB.111.134201}, with the crucial difference that the projector is constructed from left and right eigenvectors.

Throughout this work, \(m\) denotes an open-boundary energy sector adiabatically connected to the \(m\)-th non-Bloch band. In the numerical examples below, we focus on the sector with negative real energy, denoted by \(m=-\).

\begin{figure}[hbpt]
	\includegraphics[width=8.6 cm]{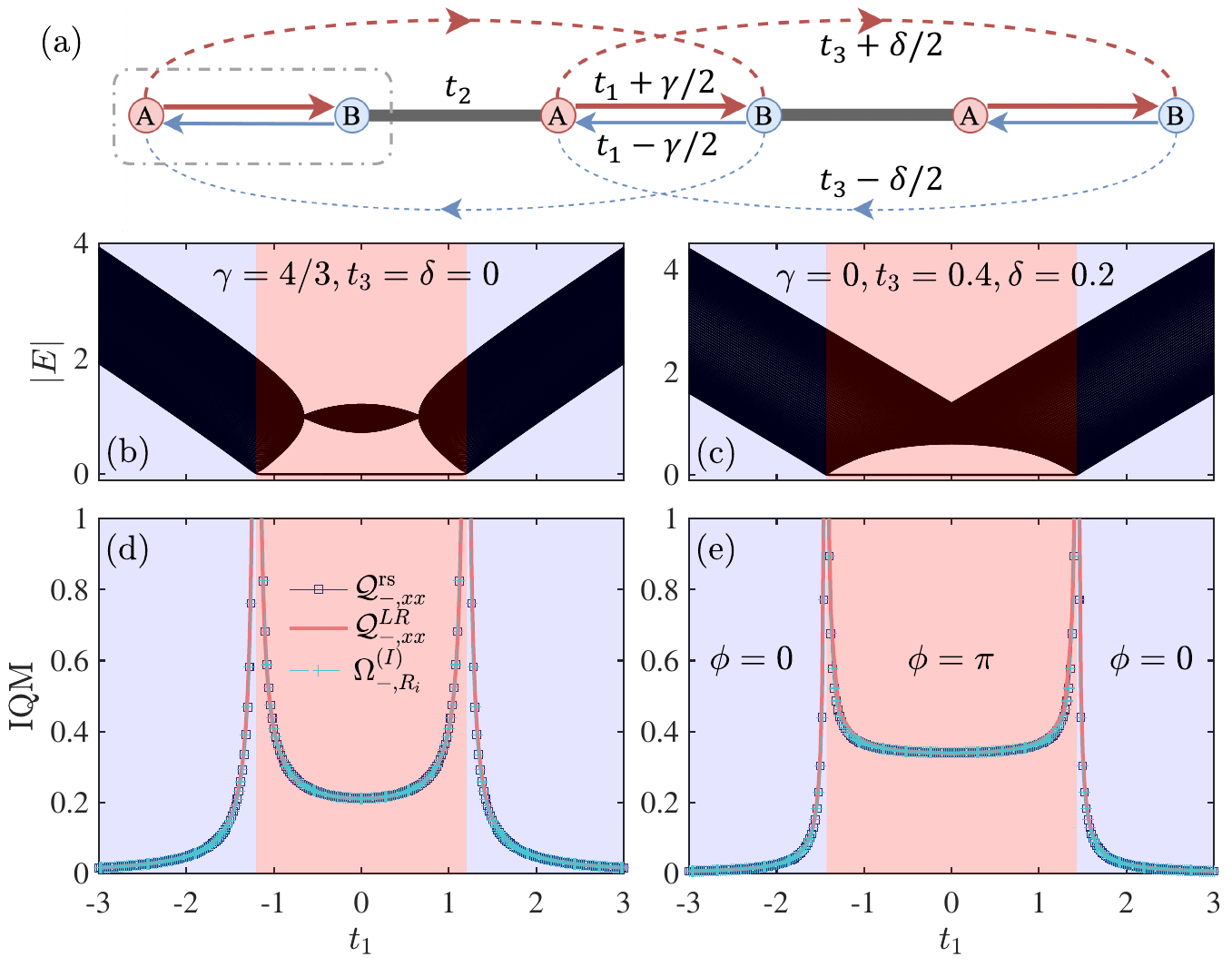}
	\caption{
	(a) Schematic of the non-Hermitian SSH model with nonreciprocal hopping. Each unit cell, indicated by the dashed box, contains two sublattices, A and B.
	(b) and (c) Absolute values of the energy spectrum as functions of \(t_1\) for different parameter choices.
	(d) and (e) show the corresponding real-space IQM (\(\mathcal{Q}^{\rm rs}_{-,xx}\)), non-Bloch IQM (\(\mathcal{Q}^{LR}_{-,xx}\)), and projected-position Wannier function spread (\(\Omega^{(I)}_{-,R_i}\)) as a function of \(t_1\) for the cases shown in (b) and (c), respectively.
	In (b) and (c), \(t_2=1\). The red-shaded region denotes the topological phase, characterized by the Berry phase \(\phi=\pi\) [see Eq.~(\ref{eqWcenter})], whereas the blue-shaded region denotes the topologically trivial phase with \(\phi=0\).
	}\label{fig1}
\end{figure}

We apply Eq.~(\ref{eqRIQGT}) to the prototypical non-Hermitian Su-Schrieffer-Heeger (SSH) model~\cite{PhysRevLett.121.086803, PhysRevLett.123.246801, PhysRevA.97.052115, PhysRevB.97.045106} shown in Fig.~\ref{fig1}(a). The model contains intracell nonreciprocal hoppings \(t_1\pm\gamma/2\), intercell reciprocal hopping \(t_2\), and long-range nonreciprocal hoppings \(t_3\pm\delta/2\). As \(t_1\) is varied, the open-boundary spectrum displays a trivial phase, shown by the blue-shaded region, and a topological phase with zero-energy modes, shown by the red-shaded region, as predicted by non-Bloch theory~\cite{PhysRevLett.121.086803, PhysRevLett.123.246801} [see Figs.~\ref{fig1}(b) and \ref{fig1}(c), which correspond to the simple case where the GBZ is a circle and the more general case where the GBZ is not a circle, respectively].

Figures~\ref{fig1}(d) and \ref{fig1}(e) show that the real-space IQM detects this open-boundary topological transition. The quantity \(\mathcal{Q}^{\rm rs}_{-,xx}\) is relatively large in the topological regime, small in the trivial regime, and begins to diverge near the gap-closing point. Thus, the real-space non-Hermitian IQM captures the distinct quantum-geometric characteristics of different open-boundary phases. Below we show that this behavior has a direct localization interpretation in terms of non-Bloch Wannier functions.

\textit{Equivalence between real-space and non-Bloch quantum geometry.---}
We now establish the connection between the real-space IQM and non-Bloch band theory. Under open-boundary conditions, the eigenstates of a non-Hermitian system are composed of right and left non-Bloch waves~\cite{PhysRevLett.121.086803, ZhangXintong_2024, PhysRevB.100.035102, ZhangKai2025,SM},
\[
|\psi^R_{m,\beta}\rangle
=
\frac{1}{\sqrt{N}}\beta^x|u^R_{m,\beta}\rangle,
\quad
\langle\psi^L_{m,\beta}|
=
\frac{1}{\sqrt{N}}\langle u^L_{m,\beta}|\beta^{-x},
\]
which satisfy
\(\langle \psi^{L}_{m,\beta} | \psi^{R}_{m,\beta} \rangle=1\).
Here, \(|u^R_{m,\beta}\rangle\) and \(\langle u^L_{m,\beta}|\) are right and left eigenvectors of the non-Bloch Hamiltonian \(\mathcal{H}_\beta\),
\(
\mathcal{H}_\beta |u^R_{m,\beta}\rangle
=
E_{m,\beta}|u^R_{m,\beta}\rangle,
\ 
\langle u^L_{m,\beta}|\mathcal{H}_\beta
=
E_{m,\beta}\langle u^L_{m,\beta}|,
\)
with
\(\langle u^L_{m,\beta}|u^R_{n,\beta}\rangle=\delta_{mn}\).
The index \(m\) labels non-Bloch bands, while \(\beta\) parametrizes the GBZ. A general non-Bloch Hamiltonian \(\mathcal{H}_\beta\) can be readily obtained from the Bloch Hamiltonian \(H(k)\) by replacing \(e^{ik}\) by \(\beta\), namely \(H(k)\to\mathcal{H}_\beta\) \cite{PhysRevLett.121.086803}.

A non-Bloch wave is therefore a product of a unit-cell-periodic component, determined by \(\mathcal{H}_\beta\), and an exponential factor, \(\beta^{x}\) or \(\beta^{-x}\).  In general,
\(\beta=|\beta(k)|e^{ik}\)
is complex, and its modulus depends on its argument (denoted by $k$) . When the GBZ reduces to the conventional Brillouin zone, non-Bloch waves reduce to ordinary Bloch waves. Open-boundary eigenstates are linear combinations of non-Bloch waves. For example, a right eigenstate may be written as
\(
|\Psi^R\rangle
=
c_1|\psi^R_{m,\beta_1}\rangle
+
c_2|\psi^R_{m,\beta_2}\rangle,
\)
where \(\beta_1\) and \(\beta_2\) typically form a complex-conjugate pair, \(\beta_2^*=\beta_1\), so that the wave function satisfies open-boundary conditions through destructive interference at the boundaries. Left eigenstates can be constructed analogously from left non-Bloch waves.

Because non-Bloch waves encode the eigenfunctions of the open-boundary real-space Hamiltonian, they provide a natural basis for quantum geometry. The projector onto the \(m\)-th sector can be written as
\begin{align}\label{eqProjnonBloch}
\hat{P}_{m}
=
\frac{N}{2\pi}
\int
\frac{d\beta}{i\beta}
|\psi^R_{m,\beta}\rangle
\langle\psi^L_{m,\beta}|.
\end{align}
The integration measure \(d\beta/(i\beta)\) is essential in the non-Hermitian case because it ensures the completeness relation
\(\sum_m \hat{P}_m=\mathbf{I}\).
In the Hermitian limit, \(\beta\to e^{ik}\) and \(d\beta/(i\beta)\to dk\), so the projector reduces to its conventional Hermitian counterpart. Substituting the non-Bloch projector in Eq.~(\ref{eqProjnonBloch}) into Eq.~(\ref{eqRIQGT}), we obtain the central equivalence relation of this work [see SM \cite{SM}]:
\begin{align}\label{eqQMxx}
\begin{split}
	\mathcal{Q}^{\rm rs}_{m,xx}
	&=
	\frac{1}{2\pi}
	\int_{\rm GBZ}
	\chi^{LR}_{m,xx}(k)dk\equiv \mathcal{Q}^{LR}_{m,xx},
	\\
	\chi^{LR}_{m,xx}
	&=
	\langle \partial_{k} u_{m,\beta}^L|
	\left[
	\mathbf{I}
	-
	|u_{m,\beta}^R\rangle
	\langle u_{m,\beta}^L|
	\right]
	|\partial_{k} u_{m,\beta}^R\rangle,
\end{split}
\end{align}
where \(\chi^{LR}_{m,xx}\) is the left-right quantum metric tensor defined using the non-Bloch Hamiltonian, and  its integral over the GBZ defines the non-Bloch IQM \(\mathcal{Q}^{LR}_{m,xx}\).  \(\chi^{LR}_{m,xx}\) is invariant under the GL(1,\(\mathbb{C}\)) gauge transformation
\(
|u^R_{m,\beta}\rangle
\rightarrow
z(\beta)|u^R_{m,\beta}\rangle,
\quad
\langle u^L_{m,\beta}|
\rightarrow
z(\beta)^{-1}\langle u^L_{m,\beta}|,
\)
where \(z(\beta)\) is an arbitrary nonzero complex function preserving biorthonormality~\cite{NHQG_Kawabata}. The identity \(\mathcal{Q}^{\rm rs}_{m,xx}
\equiv \mathcal{Q}^{LR}_{m,xx}\) in Eq.~(\ref{eqQMxx}) shows that quantum geometry under open-boundary conditions can be formulated equivalently in real-space and in non-Bloch momentum space. Numerically, we find perfect agreement between \(Q^{LR}_{-,xx}\) and \(Q^{\rm rs}_{-,xx}\) [Figs.~\ref{fig1}(d) and \ref{fig1}(e)], confirming the equivalence.

\textit{Non-Bloch Wannier functions and localization.---}
We next clarify the physical meaning of the non-Bloch IQM. In Hermitian systems, the IQM gives the gauge-invariant part of the Wannier spread functional and therefore sets a lower bound on Wannier function localization~\cite{PhysRevB.56.12847, RevModPhys.84.1419}. We now show that an analogous relation holds in non-Hermitian systems, provided that Wannier functions are constructed from non-Bloch waves.

We define the right and left non-Bloch Wannier functions as
\begin{align}\label{eqWan}
\begin{split}
|w^R_{m,R_i}\rangle
&=
\sqrt{N}
\int
\frac{d\beta}{2\pi i\beta}
\beta^{-R_i}
|\psi^R_{m,\beta}\rangle,
\\
\langle w^L_{m,R_i}|
&=
\sqrt{N}
\int
\frac{d\beta}{2\pi i\beta}
\langle\psi^L_{m,\beta}|
\beta^{R_i},
\end{split}
\end{align}
where \(R_i\) denotes the position of the \(i\)-th unit cell. These functions satisfy the biorthogonality condition
\(
\langle w^L_{m,R_i'}|w^R_{n,R_i}\rangle
=
\delta_{mn}\delta_{R_i',R_i},
\)
and inherit lattice translational symmetry (\( w^{L/R}_{m,R_i}(x-R_i^\prime)=w^{L/R}_{m,R_i+R_i^\prime}(x)\), see SM \cite{SM}). When the GBZ reduces to the conventional Brillouin zone, \(\beta\to e^{ik}\) and \(d\beta/(i\beta)\to dk\), Eq.~(\ref{eqWan}) reduces to the usual Wannier functions of Hermitian systems.

To illustrate how non-Hermiticity reshapes Wannier functions, consider the model in Fig.~\ref{fig1}(a) with \(t_3=\delta=0\), corresponding to the parameters in Fig.~\ref{fig1}(b). In this case, the GBZ is a circle,
\[
\beta=re^{ik},
\qquad
r=
\sqrt{
\left|
\frac{t_1-\gamma/2}{t_1+\gamma/2}
\right|
},
\]
where \(r\neq1\) for \(\gamma\neq0\) and is independent of \(k\)~\cite{PhysRevLett.121.086803}. Equation~(\ref{eqWan}) then becomes
\begin{align}\label{eqWanV2}
\begin{split}
|w^R_{m,R_i}\rangle
&=
r^{x-R_i}
\left[
\frac{\sqrt{N}}{2\pi}
\int_{\rm GBZ}
dk\,
e^{ik(x-R_i)}
|u^R_{m,\beta}\rangle
\right],
\\
\langle w^L_{m,R_i}|
&=
\left[
\frac{\sqrt{N}}{2\pi}
\int_{\rm GBZ}
dk\,
\langle u^L_{m,\beta}|
e^{-ik(x-R_i)}
\right]
r^{-(x-R_i)}.
\end{split}
\end{align}
The expressions in square brackets have the same form as Wannier functions constructed from Bloch waves in Hermitian systems and are spatially localized. Multiplication by the real exponential factor \(r^{\pm(x-R_i)}\) preserves localization but makes the decay asymmetric about the center \(R_i\): the right Wannier function decays more slowly on one side and more rapidly on the other, while the left Wannier function is modified in the opposite way [see SM~\cite{SM}]. Thus, the skin effect directly reshapes the spatial profile of non-Bloch Wannier functions. In more general cases where the GBZ is not a circle, as the parameters in Fig.~\ref{fig1}(c), non-Hermiticity similarly modifies the Wannier function distribution.

The Wannier center defined by a pair of left and right non-Bloch Wannier functions is given by
\begin{align}\label{eqWcenter}
\begin{split}
\bar{x}^m_{R_i}
=
\langle w^L_{m,R_i}|\hat{x}|w^R_{m,R_i}\rangle
=
R_i+\frac{\phi}{2\pi},
\end{split}
\end{align}
where
\(
\phi
=
\int_{\rm GBZ}
i
\langle u^L_{m,\beta}|
\partial_k u^R_{m,\beta}
\rangle dk
\) 
is the non-Bloch Berry phase. Thus, the non-Bloch Berry phase determines the Wannier center. In the topological phase, \(\phi=\pi\), and
\(\bar{x}^m_{R_i}=R_i+1/2\), which lies between neighboring unit cells. In the trivial phase, \(\phi=0\), and
\(\bar{x}^m_{R_i}=R_i\), which lies within a unit cell.

The spread functional of a pair of left and right non-Bloch Wannier functions in the \(m\)-th band is
\[
\Omega_{m,R_i}
=
\langle w^L_{m,R_i}|
(\hat{x}-\bar{x}^m_{R_i})^2
|w^R_{m,R_i}\rangle.
\]
Because the non-Bloch Wannier functions inherit translational symmetry,
\(\Omega_{m,R_i}\) is independent of \(R_i\). Therefore, we have the relation \(
	\Omega_m
	=
	\frac{1}{N}
	\sum_{R_i}
	\Omega_{m,R_i}
	=
	\Omega_{m,R_i}\). The quantity \(\Omega_m\) can be decomposed into a gauge-invariant part \(\Omega_m^I\) and a gauge-dependent part \(\Omega_m^D\)~\cite{PhysRevB.56.12847, RevModPhys.84.1419},
\(
\Omega_m=\Omega_m^I+\Omega_m^D,
\)
with
\begin{align}\label{eqSpread}
\begin{split}
\Omega_m^I
&=
\frac{1}{N}
{\rm Tr}
\left\{
\hat{W}_m\hat{x}
(\mathbf{I}-\hat{W}_m)
\hat{x}
\right\},
\\
\Omega_m^D
&=
\frac{1}{N}
\sum_{R_i}
\sum_{R_i'(\neq R_i)}
\langle w^L_{m,R_i}|
\hat{x}
|w^R_{m,R_i'}\rangle
\langle w^L_{m,R_i'}|
\hat{x}
|w^R_{m,R_i}\rangle,
\nonumber
\end{split}
\end{align}
where $
\hat{W}_m
=
\sum_{R_i}
|w^R_{m,R_i}\rangle
\langle w^L_{m,R_i}|
$
is the projector onto the \(m\)-th band constructed from the non-Bloch Wannier functions. One can readily prove that \(\hat{W}_m=\hat{P}_m\), and that the real-space IQM in Eq.~(\ref{eqRIQGT}) can be expressed in an equivalent form as \(\mathcal{Q}^{\rm rs}_{m,xx}=
\frac{1}{N}{\rm Tr}
\left\{
\hat{P}_m\hat{x}(\mathbf{I}-\hat{P}_m)\hat{x}
\right\}\) [see SM \cite{SM}]. Hence, we immediately see that
\begin{align}
\Omega_m^I
\equiv
\mathcal{Q}^{LR}_{m,xx}
\equiv
\mathcal{Q}^{\rm rs}_{m,xx}.
\end{align}
This is the second central result of this work: the non-Bloch IQM is exactly the gauge-invariant part of the spread functional of localized non-Bloch Wannier functions. Therefore, the non-Bloch  IQM directly characterizes their localization properties.

\begin{figure}[hbpt]
	\includegraphics[width=7.6 cm]{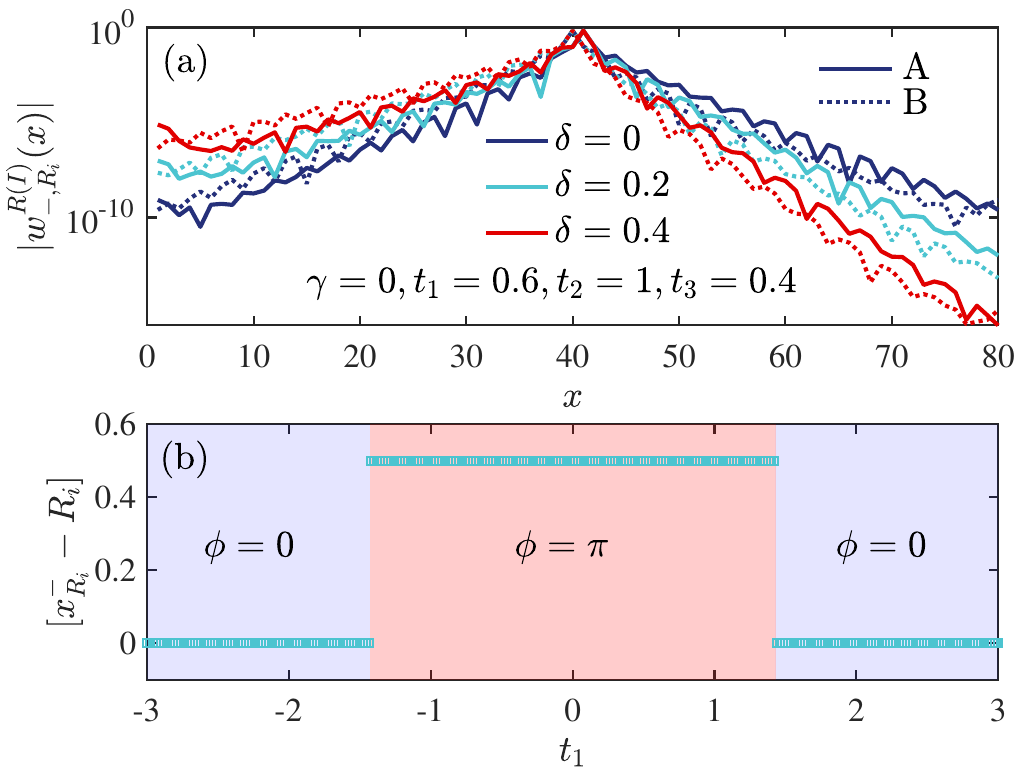}
	\caption{
	(a) Spatial profiles of \(|w^{R(I)}_{-,R_i}\rangle\) for different values of \(\delta\), shown on a logarithmic scale along the \(y\)-axis. Solid and dashed lines denote the distributions on the two sublattices, respectively. \(\delta=0\) corresponds to the Hermitian case.
	(b) Wannier center \([x_{R_i}^- -R_i]\) as a function of \(t_1\). The parameters are the same as in Fig.~\ref{fig1}(c).
	}\label{fig2}
\end{figure}

\textit{Projected-position Wannier functions and numerical verification.---}
We numerically verify the relation between the IQM and Wannier function localization by constructing a set of non-Bloch Wannier functions for which the gauge-dependent part of the spread functional is zero. We refer to them as projected-position Wannier functions, because they are obtained by diagonalizing the projected position operator	\(\hat{P}_m\hat{x}\hat{P}_m\)~\cite{PhysRevB.26.4269, PhysRevB.56.12847}, denoted as \(|w^{R(I)}_{m,R_i}\rangle\) and \(\langle w^{L(I)}_{m,R_i}|\) (The superscript $(I)$ indicates that only the gauge-independent part of its spread functional is non-zero). They satisfy
\begin{align}\label{eqWIn}
\begin{split}
\hat{P}_m \hat{x} \hat{P}_m
|w^{R(I)}_{m,R_i}\rangle
&=
\bar{x}^m_{R_i}
|w^{R(I)}_{m,R_i}\rangle,
\\
\langle w^{L(I)}_{m,R_i}|
\hat{P}_m \hat{x} \hat{P}_m
&=
\langle w^{L(I)}_{m,R_i}|
\bar{x}^m_{R_i},
\end{split}
\end{align}
where the eigenvalue
\(
\bar{x}^m_{R_i}
=
\langle w^{L(I)}_{m,R_i}|
\hat{x}
|w^{R(I)}_{m,R_i}\rangle
\)
is the corresponding Wannier center. For these projected-position Wannier functions, the gauge-dependent part of the spread vanishes, and hence
\(
\Omega_{m,R_i}^{(I)}
=
\langle w^{L(I)}_{m,R_i}|
(\hat{x}-\bar{x}^m_{R_i})^2
|w^{R(I)}_{m,R_i}\rangle
=
\Omega_m^I
=
\mathcal{Q}^{\rm rs}_{m,xx}
=
\mathcal{Q}^{LR}_{m,xx}.
\)

Figure~\ref{fig2}(a) shows the spatial profiles of the projected-position Wannier functions. As the non-Hermitian parameter \(\delta\) increases, the decay of
\(|w^{R(I)}_{m,R_i}\rangle\)
becomes faster on the right side of its peak and slower on the left side, while
\(\langle w^{L(I)}_{m,R_i}|\)
exhibits the opposite behavior [see SM \cite{SM}]. This agrees with the previously predicted modification of non-Bloch Wannier functions by the non-Hermitian skin effect [see Eq. (\ref{eqWanV2})]. Figure~\ref{fig2}(b) verifies the relation between the Wannier center and the non-Bloch Berry phase in Eq.~(\ref{eqWcenter}). We also confirm the equality
\(
\Omega_{m,R_i}^{(I)}
=
\mathcal{Q}^{\rm rs}_{m,xx}
=
\mathcal{Q}^{LR}_{m,xx}
\), 
as shown in Figs.~\ref{fig1}(d) and \ref{fig1}(e).

\textit{Complex IQM in biorthogonal geometry.---}
Unlike in Hermitian systems, the non-Hermitian IQM and the corresponding Wannier spread need not be real and nonnegative. This originates from the biorthogonal structure of left and right Wannier functions. To see this, we define
\[
f(x)
=
[w^{L(I)}_{m,R_i}(x)]^*
w^{R(I)}_{m,R_i}(x)
\]
as the distribution of the product of left and right projected-position Wannier functions.
In Hermitian systems, left and right Wannier functions coincide, so \(f(x)\) is a nonnegative real function. The spread functional then directly measures the spatial localization of the Wannier function, and the IQM gives its gauge-invariant lower bound. In non-Hermitian systems, however, \(f(x)\) is generally complex. Consequently, the projected-position Wannier function spread
\(\Omega_{m,R_i}^{(I)}\)
and the IQM
\(\mathcal{Q}^{LR}_{m,xx}\)
can also become complex.

\begin{figure}[hbpt]
  \includegraphics[width=8.6 cm]{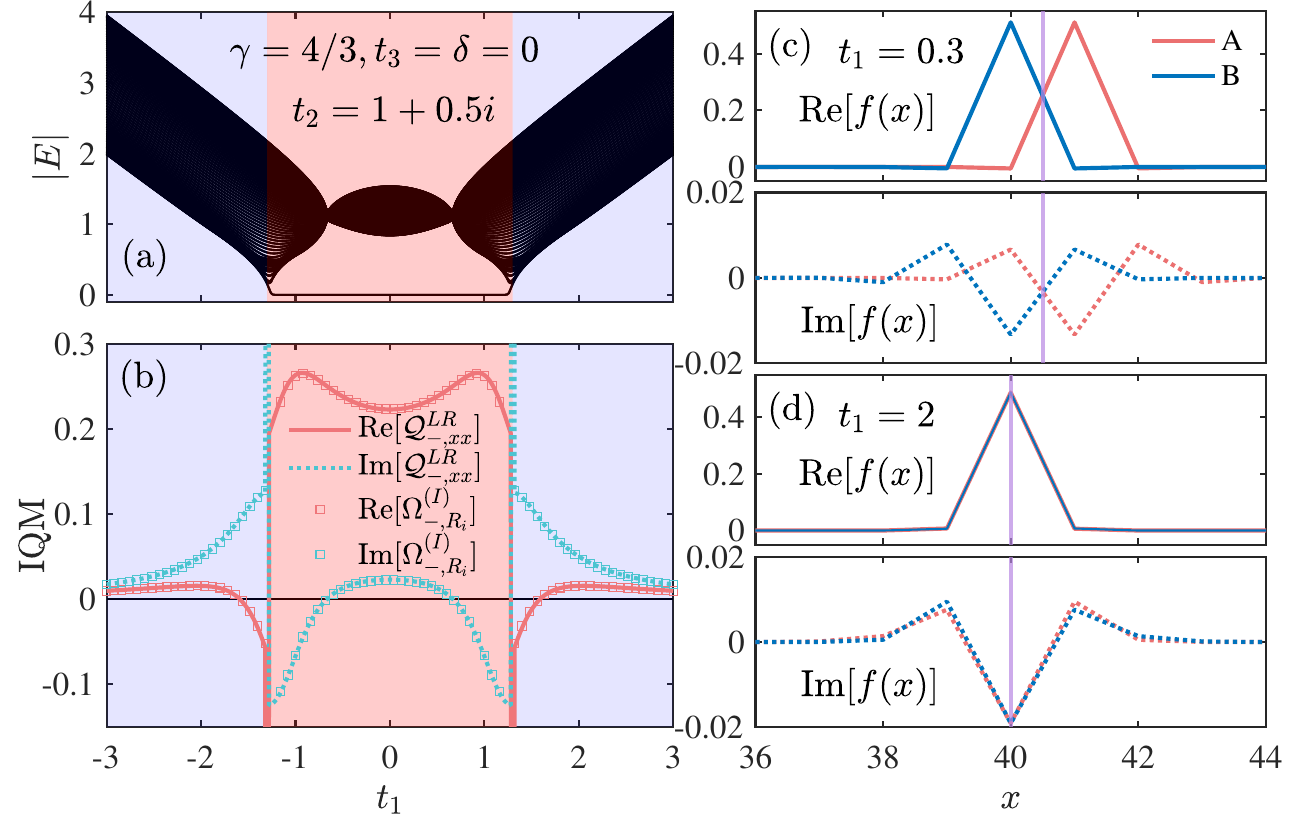}
  \caption{
  (a) and (b) Absolute value of the energy spectrum and IQM as functions of \(t_1\), respectively, for the same set of parameters. The red and blue shaded regions indicate the topological and trivial phases, respectively.
  (c) Real part, upper panel, and imaginary part, lower panel, of \(f(x)\) at \(t_1=0.3\), corresponding to the topological phase.
  (d) Same as (c), but for \(t_1=2\), corresponding to the trivial phase.
  The vertical lines in (c) and (d) mark the Wannier centers \(\bar{x}^m_{R_i}
  =
  \langle w^{L(I)}_{m,R_i}|\hat{x}|w^{R(I)}_{m,R_i}\rangle\)  defined by a pair of left and right projected-position Wannier functions.
  }\label{fig3}
\end{figure}

The projected-position Wannier function spread can be written in terms of \(f(x)\) as
\[
\Omega_{m,R_i}^{(I)}
=
\sum_x
f(x)
(x-\bar{x}^m_{R_i})^2.
\]
We decompose \(f(x)\) into sign-resolved real and imaginary components,
\[
f(x)
=
f^{\rm Re}_{+}(x)
+
f^{\rm Re}_{-}(x)
+
i
[
f^{\rm Im}_{+}(x)
+
f^{\rm Im}_{-}(x)
],
\]
where \(f^{\rm Re}_{+}\) and \(f^{\rm Re}_{-}\) denote the positive and negative real components, while \(f^{\rm Im}_{+}\) and \(f^{\rm Im}_{-}\) denote the positive and negative imaginary components. For each component, we define
\[
\Omega^{\rm Re}_{\pm}
=
\sum_x
f^{\rm Re}_{\pm}(x)
(x-\bar{x}^m_{R_i})^2,
\ \
\Omega^{\rm Im}_{\pm}
=
\sum_x
f^{\rm Im}_{\pm}(x)
(x-\bar{x}^m_{R_i})^2.
\]
Because the sign remains identical within each part,
\(|\Omega^{\rm Re}_{\pm}|\) and
\(|\Omega^{\rm Im}_{\pm}|\)
quantify the localization of the corresponding component relative to the Wannier center \(\bar{x}^m_{R_i}\). Thus,
\begin{align}\label{eqOmega_ReIm}
\begin{split}
{\rm Re}
\left[
\Omega_{m,R_i}^{(I)}
\right]
&=
|\Omega^{\rm Re}_{+}|
-
|\Omega^{\rm Re}_{-}|,
\\
{\rm Im}
\left[
\Omega_{m,R_i}^{(I)}
\right]
&=
|\Omega^{\rm Im}_{+}|
-
|\Omega^{\rm Im}_{-}|.
\end{split}
\end{align}
Accordingly, the real part of the IQM measures the difference between the localization of the positive and negative real components of \(f(x)\), while the imaginary part measures the corresponding difference between the positive and negative imaginary components.

Although \(f(x)\) can generally be complex, the biorthogonal normalization condition
\(
\langle w^{L(I)}_{m,R_i}|w^{R(I)}_{m,R_i}\rangle
=
\sum_x f(x)
=
1
\)
strongly constrains its distribution. In the examples considered below, the localization is governed primarily by the real part of \(f(x)\), while the imaginary component is small and arises from the biorthogonal structure.

For the parameters in Fig.~\ref{fig3}(a), varying \(t_1\) drives a transition between the topological phase and the trivial phase. The corresponding IQM becomes complex [Fig.~\ref{fig3}(b)]. Figures~\ref{fig3}(c) and \ref{fig3}(d) show \(f(x)\) in the topological phase \((t_1=0.3)\) and in the trivial phase \((t_1=2)\), respectively. The real part of \(f(x)\) clearly distinguishes the two phases. In the topological phase, it is more spatially extended, leading to a relatively large positive value of
\({\rm Re}[\Omega_{m,R_i}^{(I)}]\). In the trivial phase, it is more localized, yielding a smaller value of
\({\rm Re}[\Omega_{m,R_i}^{(I)}]\), which can be either weakly positive or weakly negative. Thus, the real part of the IQM provides a physically meaningful measure of localization across different non-Hermitian phases.

\textit{Conclusion and discussion.---}
We have developed a quantum-geometric framework for non-Hermitian systems under open boundary conditions. By defining the IQM directly in real space and relating it to non-Bloch band theory, we established an exact equivalence between the real-space IQM and the non-Bloch IQM defined on the GBZ. We further constructed localized non-Bloch Wannier functions and showed that the non-Bloch IQM equals the gauge-invariant part of their spread functional. This establishes a direct physical meaning of non-Bloch quantum geometry: it characterizes the localization properties of Wannier functions built from skin modes.

Our results provide a quantum-geometric foundation for non-Hermitian band theory beyond conventional Bloch descriptions. They also suggest that geometric diagnostics, localization bounds, and topological characterizations can be extended to broad classes of open-boundary non-Hermitian systems where the skin effect plays an essential role.

\begin{acknowledgments}
J.S. acknowledges support from China Scholarship Council. H.G. acknowledge support from the NSFC grant No.~12574249 and the BNLCMP open research fund under Grant No.~2024BNLCMPKF023. B.-J.Y was supported by Samsung Science and Technology Foundation under project no. SSTF-BA2002-06, National Research Foundation of Korea (NRF) funded by the Korean government(MSIT), grant no. RS-2021-NR060087 and RS-2025-00562579, Global Research Development-Center (GRDC) Cooperative Hub Program through the NRF funded by the MSIT, grant no. RS-2023-00258359, Global-LAMP program of the NRF funded by the Ministry of Education, grant no. RS-2023-00301976.
\end{acknowledgments}

\bibliographystyle{apsrev4-1-etal-title_10authors}
\bibliography{ref}

\end{document}


\title{Supplemental Material for "Non-Bloch Quantum Geometry of Non-Hermitian Systems"}
\author{Junsong Sun}
\affiliation{Department of Physics and Astronomy, Seoul National University, Seoul 08826, Korea}
\affiliation{School of Physics, Beihang University,
Beijing, 100191, China}

\author{Huaiming Guo}
\email{hmguo@buaa.edu.cn}
\affiliation{School of Physics, Beihang University,
Beijing, 100191, China}

\author{Bohm-Jung Yang}
\email{bjyang@snu.ac.kr}
\affiliation{Department of Physics and Astronomy, Seoul National University, Seoul 08826, Korea}
\affiliation{Center for Theoretical Physics (CTP), Seoul National University, Seoul 08826, Korea}
\affiliation{Institute of Applied Physics, Seoul National University, Seoul 08826, Korea}
\date{\today}
\begin{abstract}

\end{abstract}
\maketitle
\renewcommand{\thefigure}{S\arabic{figure}}
\renewcommand{\theequation}{S\arabic{equation}}
\renewcommand{\thesection}{S\arabic{section}}
\setcounter{figure}{0}  
\setcounter{equation}{0} 
\numberwithin{equation}{section}

\tableofcontents

\section{Non-Bloch wave and generalized Brillion zone}

\begin{figure}[hbpt]
	\includegraphics[width=10.8 cm]{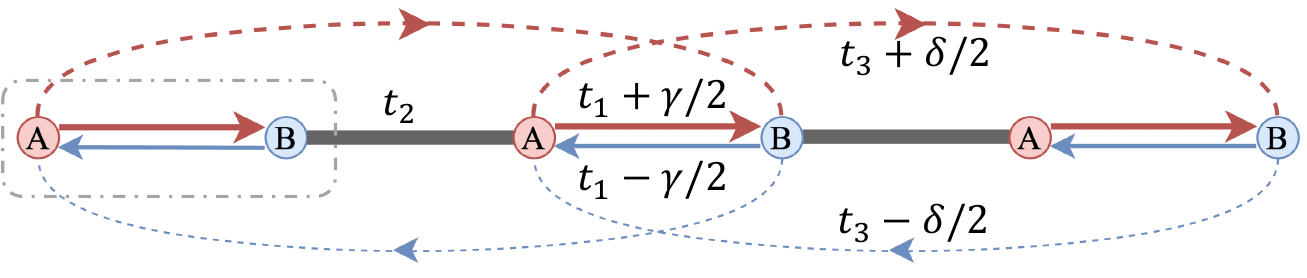}
	\caption{Schematic of the non-Hermitian SSH model with nonreciprocal hopping. Each unit cell (indicated by the dashed box) contains two sublattices, A and B.
	}\label{figS0}
\end{figure}

We now briefly review the general solution of the non-Hermitian SSH model shown in Fig.~\ref{figS0}. For simplicity, we present the analytical solution for the case \(t_3 = \delta = 0\) as an example; the more general case with \(t_3 \neq 0\) and \(\delta \neq 0\) can be solved in a similar manner. Assuming the real-space right eigenfunction to take the form
\begin{align}\label{}
\begin{split}
& |\psi^R_\beta\rangle =(\psi^R_{1,A},\psi^R_{1,B},\cdots,\psi^R_{L,A},\psi^R_{L,B})^T,\\
 &(\psi^R_{x,A},\psi^R_{x,B})=\beta^x(\phi^R_{A}, \phi^R_{B}).
 \end{split}
\end{align}
The wave function thus takes the form of a product of a unit-cell periodic function, \((\phi^R_A, \phi^R_B)\), and an exponential factor \(\beta^x\).
In the bulk, the eigenvalue equation of the real-space Hamiltonian yields
\begin{align}\label{}
\begin{split}
 &t_2 \psi^R_{x-1,B} + (t_1 + \gamma/2)\psi^R_{x,B} = E_\beta\psi^R_{x,A},\\
 &(t_1-\gamma/2) \psi^R_{x,A} + t_2\psi^R_{x+1,A} = E_\beta\psi^R_{x,B},
\end{split}
\end{align}
this implies that the unit-cell periodic component must satisfy
\begin{align}\label{eqperiod}
	\left[\begin{array}{cc}
		0 & t_1+\frac{\gamma}{2}+t_2\beta^{-1} \\
		t_1-\frac{\gamma}{2}+t_2\beta & 0
	\end{array}\right]\left[\begin{array}{c} \phi^R_A \\
		\phi^R_B
	\end{array}\right]=E_\beta\left[\begin{array}{c} \phi^R_A \\
		\phi^R_B
	\end{array}\right].
\end{align}
We observe that the matrix on the left-hand side of Eq. (\ref{eqperiod}) exactly takes the form of the non-Bloch Hamiltonian of the non-Hermitian SSH model.
A general non-Bloch Hamiltonian for non-Hermitian systems can be readily obtained from the Bloch Hamiltonian \(H(k)\) by replacing the phase factor as \(e^{ik} \rightarrow \beta\), with \(\beta = |\beta(k)| e^{ik}\) parametrizing the generalized Brillion zone(GBZ). We denote the non-Bloch Hamiltonian as \(\mathcal{H}_\beta\),
\begin{align}\label{eqHnonBloch}
	\mathcal{H}_\beta=\left[\begin{array}{cc}
			0 & t_1+\frac{\gamma}{2}+t_2\beta^{-1} \\
			t_1-\frac{\gamma}{2}+t_2\beta & 0
		\end{array}\right].
\end{align}
Its right (left) eigenvectors satisfy \(\mathcal{H}_\beta |u_{m,\beta}^R\rangle = E_{m,\beta} |u_{m,\beta}^R\rangle\) (\( \langle u_{m,\beta}^L| \mathcal{H}_\beta= E_{m,\beta} \langle u_{m,\beta}^L|\)), where the subscript \(m\) labels different bands. The left and right eigenvectors satisfy the biorthogonality condition, \( \langle u_{m,\beta}^L | u_{n,\beta}^R \rangle = \delta_{mn} \). Accordingly, its right eigenvector determines the unit-cell periodic part of the right wave function \(|\psi^R_{m,\beta}\rangle\), namely \((\phi_A^R, \phi_B^R)^T = |u_{m,\beta}^R\rangle\), with \(E_\beta=E_{m,\beta}\) (Here, we have introduced an additional subscript \(m\) to \(|\psi_\beta\rangle\) and \(E_\beta\) to label different eigenstates of the non-Bloch Hamiltonian).
This correspondence explains why the spectrum obtained from the non-Bloch Hamiltonian correctly predicts the real-space spectrum under open boundary conditions. It further indicates that the topological properties of the non-Hermitian SSH model under open boundary conditions are encoded in the periodic component of the wave function, thereby the topological invariants defined within the non-Bloch framework can correctly capture the topology under open boundaries. 

\begin{figure}[hbpt]
	\includegraphics[width=12.8 cm]{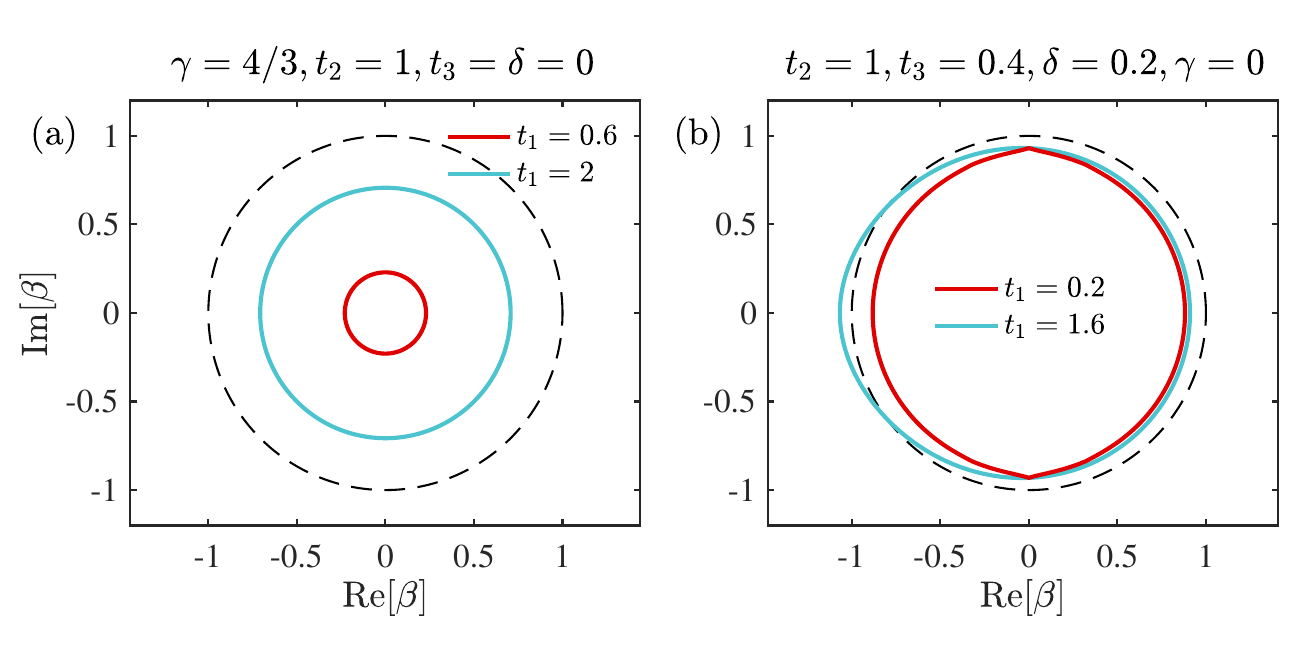}
	\caption{Generalized Brillouin zones of the non-Hermitian SSH model shown in Fig.~\ref{figS0} for different parameter settings. The black dashed line denotes the unit circle.
	}\label{figGBZ}
\end{figure}

Together with the boundary conditions, in the long-chain limit one obtains two solutions for \(\beta\), which must satisfy 
\begin{align}\label{GBZ_ssh}
|\beta_1| = |\beta_2|=\sqrt{\left|\frac{t_1-\gamma/2}{t_1+\gamma/2}\right|}.
\end{align}
Here, the modulus \( |\beta| \) being a constant, such that the corresponding GBZ forms a circle [see Fig.~\ref{figGBZ}(a)], is a special case. In general, the GBZ is not circular (for example, for the parameter set corresponding to Fig. 1(c) of the main text); instead, \( \beta = |\beta(k)| e^{ik} \), with the modulus also depending on the phase \( k \). Without loss of generality, we proceed to consider the general case \( \beta = |\beta(k)| e^{ik} \). Then the general solution of the right wave function in real space takes the form of a linear superposition of non-Bloch waves \(|\psi^R_{m,\beta}\rangle=\frac{1}{\sqrt{N}}\beta^x|u^R_{m,\beta}\rangle \), namely \(|\Psi^R\rangle =c_1|\psi^R_{m,\beta_1} \rangle +c_2|\psi^R_{m,\beta_2} \rangle\). In fact, \(\beta_1\) and \(\beta_2\) appear as a conjugate pair, \(\beta_1 = \beta_2^*\), ensuring that the real-space wave function vanishes at the boundaries. 

\section{Generalized Brillouin zone in the general case}
For a generic one-dimensional non-Hermitian tight-binding Hamiltonian with finite-range hopping up to \(m\)-th neighbors, the non-Bloch Hamiltonian \(\mathcal{H}(\beta)\) leads to a characteristic equation
\begin{align}\label{}
	\det\big[ \mathcal{H}(\beta) - E \big] = 0,
\end{align}
which is a Laurent polynomial in \(\beta\) of total degree \(2m\). Accordingly, for a given energy \(E\), this equation admits \(2m\) roots \({\beta_j(E)}\) in the complex plane. Ordering these roots by their moduli as
\begin{align}\label{eqrootsorder}
	|\beta_1| \le |\beta_2| \le \cdots \le |\beta_{2m}|,
\end{align}
the GBZ is determined by the condition
\begin{align}\label{}
	|\beta_m| = |\beta_{m+1}|.
\end{align}
This selection rule identifies the subset of solutions that contribute to physical bulk states under open boundary conditions in the thermodynamic limit. The resulting GBZ forms a closed loop in the complex \(\beta\)-plane, along which the non-Bloch Hamiltonian correctly reproduces the open-boundary spectrum and eigenstates.

Next, using the model shown in Fig.~\ref{figS0} with parameters \(\gamma = 0\), \(t_3 \neq 0\), and \(\delta \neq 0\) as an example, we illustrate the generic situation in which the GBZ solutions \(\beta\) do not form a circle. In this case, the non-Bloch Hamiltonian \(\mathcal{H}_\beta\) is obtained from the Bloch Hamiltonian by the substitution \( e^{ik} \to \beta \), 
\begin{align}\label{eqnBH_Gssh}
\mathcal{H}_\beta=\left[\begin{array}{cc}
	0 & t_1+t_2\beta^-+(t_3+\frac{\delta}{2})\beta \\
	t_1+t_2\beta+(t_3-\frac{\delta}{2})\beta^- & 0
\end{array}\right].
\end{align}
Its eigenvalues are determined by the characteristic equation \(\det\big[ \mathcal{H}(\beta) - E \big] = 0\), which requires the following equation to be satisfied
\begin{align}\label{eqChar}
\begin{split}
&w_4\beta^4+w_3\beta^3+w_2\beta^2+w_1\beta+w_0=0,\\
&w_4=2t_2(2t_3+\delta),\\
&w_3=2t_1[2(t_2+t_3)+\delta],\\
&w_2=4(t_1^2+t_2^2+t_3^2)-\delta^2-4E^2,\\
&w_1=2t_1[2(t_2+t_3)-\delta],\\
&w_0=2t_2(2t_3-\delta).
\end{split}
\end{align}
When \( w_4 \neq 0 \), Vieta's theorem requires that the four roots of the characteristic equation, \( \beta_1, \beta_2, \beta_3, \beta_4 \) (Note that the subscript \( n = 1, 2, 3, 4 \) here is only used to distinguish different roots and does not follow the ordering by modulus as in Eq.~(\ref{eqrootsorder})), satisfy
\begin{align}\label{eqVeta}
	\begin{split}
	&\beta_1+\beta_2+\beta_3+\beta_4=-\frac{w_3}{w_4},\\
	&\beta_1\beta_2\beta_3+\beta_1\beta_2\beta_4+\beta_1\beta_3\beta_4+\beta_2\beta_3\beta_4=-\frac{w_1}{w_4},\\
	&\beta_1\beta_2\beta_3\beta_4=\frac{w_0}{w_4}.
	\end{split}
\end{align}
We assume that the GBZ solutions satisfy \( |\beta_1| = |\beta_2| \), with \( \beta_1 = r(k)e^{ik} \) and \( \beta_2 = r(k)e^{-ik} \) (This means that \(\beta_1\) and \(\beta_2\) are assumed to be the two middle roots among the four when sorted in ascending order by modulus). By using equations in Eq. (\ref{eqVeta}) to eliminate \( \beta_3 \) and \( \beta_4 \), one obtains the relation between \( r(k) \) and \( k \) as
\begin{align}\label{eqrk}
	\begin{split}
	[2\cos k] r^4+ \frac{w_3}{w_4} r^3-\frac{w_1}{w_4}r-2\frac{w_0}{w_4}\cos k=0.
	\end{split}
\end{align}
This is a quartic equation in \( r \); for each \( k \), only one root corresponds to the GBZ solution. In this case, the generalized Brillouin zone is no longer a circle [see Fig.~\ref{figGBZ}(b)].

\section{Equivalence between real-space and non-Bloch IQM}
The real-space IQM in Eq. (1) of the main text can be expressed in an alternative but equivalent form. By expanding Eq. (1) of the main text and using the relation \(\hat{P}_m^2 = \hat{P}_m\), together with the cyclic invariance of the trace, we obtain
\begin{align}\label{eqRSQGTequival}
	\begin{split}
		\mathcal{Q}^{\rm rs}_{m,xx}&=\frac{1}{N}{\rm Tr}\left\{\hat{P}_m\hat{x}^2\hat{P}_m-\hat{P}_m\hat{x}\hat{P}_m\hat{x}\hat{P}_m\right\}=\frac{1}{N}{\rm Tr}\left\{\hat{P}_m\hat{x}(\mathbf{I}-\hat{P}_m)\hat{x}\right\}.
	\end{split}
\end{align}

Next, we provide the derivations of several key equations presented in the main text. For convenience in the subsequent derivations, we first introduce an important identity from complex analysis that will be frequently used below:
\begin{align}\label{eqsumseries}
	\oint_{\mathcal{C}} \frac{dz}{2\pi i z} f(z)\sum_{x}\left(\frac{z'}{z}\right)^x = f(z'),
\end{align}
where \( \oint_{\mathcal{C}} \) denotes the contour integral over a closed loop \( \mathcal{C} \) in the complex plane, and \( f(z) \) is a complex function analytic on and inside \( \mathcal{C} \). The above identity can be readily established. Since \( f(z) \) is analytic on and inside the contour \( \mathcal{C} \), it admits a Laurent series expansion,
\begin{align}
	f(z) = \sum_{x=-\infty}^{\infty} f_x z^x,\ \ f_x = \oint_{\mathcal{C}} \frac{dz}{2\pi i z} f(z) z^{-x}.
\end{align}
Using this expansion, we evaluate
\begin{align}
	\begin{split}
		&\oint_{\mathcal{C}} \frac{dz}{2\pi i z} f(z)\sum_{x}\left(\frac{z'}{z}\right)^x
		= \sum_{x} z'^x \oint_{\mathcal{C}} \frac{dz}{2\pi i z}f(z)z^{-x}=\sum_{x} z'^x f_x=f(z^\prime).
	\end{split}
\end{align}


First, we prove the completeness of the projector  \(\hat{P}_{m} =\frac{N}{2\pi}\int \frac{d\beta}{ i\beta}|\psi^R_{m,\beta}\rangle\langle\psi^L_{m,\beta}|\).

\begin{align}
	\begin{split}
		\sum_m \hat{P}_{m}=\int \frac{d\beta}{2\pi i\beta}\sum_{x,x^\prime}\left[\sum_m |u^R_{m,\beta}\rangle\langle u^L_{m,\beta}|\right]\otimes |x\rangle\langle x^\prime| \beta^{x-x^\prime}=\sum_x \mathbf{I}\otimes |x\rangle\langle x|=\mathbf{I}_{N\times N},
	\end{split}
\end{align}
where we have used the relation \(=\int \frac{d\beta}{2\pi i\beta}\beta^{x-x^\prime}=\delta_{x,x^\prime}\) and \(\sum_m |u^R_{m,\beta}\rangle\langle u^L_{m,\beta}|=\mathbf{I} \).

Now we present a detailed derivation of the equivalence between the real-space IQM and the non-Bloch IQM, given by
\begin{align}
	\begin{split}
		\mathcal{Q}^{\rm rs}_{m,xx}
		&=
		\frac{1}{N}{\rm Tr}
		\left\{
		\hat{P}_m\hat{x}(\mathbf{I}-\hat{P}_m)\hat{x}
		\right\}
		=
		\frac{1}{2\pi}
		\int_{\rm GBZ}
		\chi^{LR}_{m,xx}(k)dk=\mathcal{Q}^{LR}_{m,xx},
		\\
		\chi^{LR}_{m,xx}
		&=
		\langle \partial_{k} u_{m,\beta}^L|
		\left[
		\mathbf{I}
		-
		|u_{m,\beta}^R\rangle
		\langle u_{m,\beta}^L|
		\right]
		|\partial_{k} u_{m,\beta}^R\rangle.
	\end{split}
\end{align}
Substituting the projector constructed from non-Bloch states,  \(\hat{P}_{m} =\frac{N}{2\pi}\int \frac{d\beta}{ i\beta}|\psi^R_{m,\beta}\rangle\langle\psi^L_{m,\beta}|\), into Eq. (\ref{eqRSQGTequival}), we obtain
\begin{align}\label{}
	\begin{split}
		\mathcal{Q}^{\rm rs}_{m,xx}&=\frac{1}{N}{\rm Tr}\left\{\hat{P}_m\hat{x}(1-\hat{P}_m)\hat{x}\right\}=\frac{1}{N}\sum_{n\neq m}{\rm Tr}\left\{\hat{P}_m\hat{x}\hat{P}_n\hat{x}\hat{P}_m\right\}\\
		&=\frac{1}{2\pi}\left[\frac{N}{2\pi}\right]^2\sum_{n\neq m}\int \frac{d{\beta}}{i{\beta}}\frac{d{\eta}}{i{\eta}}\frac{d{\xi}}{i{\xi}} {\rm Tr}\left\{|\psi^R_{m, {\beta}}\rangle \langle\psi^L_{m, {\beta}}|x|\psi^R_{n, {\eta}}\rangle \langle\psi^L_{n, {\eta}}|x|\psi^R_{m,{\xi}}\rangle \langle\psi^L_{m,{\xi}}|\right\}\\
		&=\left[\frac{N}{2\pi}\right]^2\sum_{n\neq m}\int \frac{d{\beta}}{i{\beta}}\frac{d{\eta}}{i{\eta}}\frac{d{\xi}}{2\pi i{\xi}} \langle\psi^L_{m, {\beta}}|x|\psi^R_{n, {\eta}}\rangle \langle\psi^L_{n, {\eta}}|x|\psi^R_{m,{\xi}}\rangle \langle\psi^L_{m,{\xi}}|\psi^R_{m, {\beta}}\rangle.
	\end{split}
\end{align}
Substituting \(
\langle \psi^L_{m,\xi} | \psi^R_{m,\beta} \rangle = \frac{1}{N}\langle u^L_{m,\xi} | u^R_{m,\beta} \rangle \sum_x \left(\frac{\beta}{\xi}\right)^x\)
into the above expression, and then using Eq. (\ref{eqsumseries}) (the GBZ corresponds to a closed loop \( \mathcal{C} \)), we obtain
\begin{align}\label{eq14}
	\begin{split}
		\mathcal{Q}^{\rm rs}_{m,xx}
		&=N\sum_{n\neq m}\int \frac{d{\beta}}{2\pi i{\beta}}\frac{d{\eta}}{2\pi i{\eta}} \langle\psi^L_{m, {\beta}}|x|\psi^R_{n, {\eta}}\rangle \langle\psi^L_{n, {\eta}}|x|\psi^R_{m,{\beta}}\rangle.
	\end{split}
\end{align}
Next, we analyze \(\langle\psi^L_{n, {\eta}}|x|\psi^R_{m, {\beta}}\rangle=\frac{1}{N}\sum_{x}  u^{L*}_{n,{\eta}}\eta^{-x}x u^R_{m, {\beta}}\beta^{x}=\frac{\beta}{N\partial_k \beta}\sum_{x}  u^{L*}_{n,{\eta}}\eta^{-x}\left[\partial_k \left(u^R_{m, {\beta}}\beta^{x}\right)-\beta^{x}\partial_k u^R_{m, \beta} \right]\). Therefore
\begin{align}\label{eq15}
	\begin{split}
		\langle\psi^L_{n, {\eta}}|x |\psi^R_{m, {\beta}}\rangle=\frac{\beta}{N\partial_k \beta}\partial_{k}\left[\langle u^{L}_{n,{\eta}}| u^{R}_{m,{\beta}}\rangle\sum_{x}  \left(\frac{\beta}{\eta}\right)^{x}\right]
		-\frac{\beta}{N\partial_k \beta}\langle u^{L}_{n,{\eta}}| \partial_k u^{R}_{m,{\beta}}\rangle\sum_{x}  \left(\frac{\beta}{\eta}\right)^{x},
	\end{split}
\end{align}
where \( k \) corresponds to the phase (argument) of \( \beta \), with \( \beta = |\beta(k)| e^{ik} \).
Substituting Eq. (\ref{eq15}) into Eq. (\ref{eq14}), we then have
\begin{align}\label{eq16}
	\begin{split}
		\mathcal{Q}^{\rm rs}_{m,xx}
		&=-\sum_{n\neq m}\int \frac{dk}{2\pi i} \langle\psi^L_{m, {\beta}}|x|\psi^R_{n, {\beta}}\rangle \langle u^{L}_{n,{\beta}}| \partial_k u^{R}_{m,{\beta}}\rangle=-\sum_{n\neq m}\int \frac{dk}{2\pi i} \left[\frac{1}{N}\sum_x u^{L*}_{m,\beta}e^{-ikx}xu^R_{n,\beta}e^{ikx}\right] \langle u^{L}_{n,{\beta}}| \partial_k u^{R}_{m,{\beta}}\rangle\\
		&=\sum_{n\neq m}\int \frac{dk}{2\pi}  \langle \partial_k u^{L}_{m,{\beta}}| u^{R}_{n,{\beta}}\rangle \langle u^{L}_{n,{\beta}}| \partial_k u^{R}_{m,{\beta}}\rangle=\frac{1}{2\pi}\int \langle \partial_{k} u_{m,\beta}^L| \left[ \mathbf{I}-| u_{m,\beta}^R \rangle\langle u_{m,\beta}^L| \right]|\partial_{k} u_{m,\beta}^R \rangle dk.
	\end{split}
\end{align}

\section{Properties of the non-Bloch Wannier functions}
As shown in the main text, the non-Bloch Wannier function defined in terms of non-Bloch waves is given by
\begin{align}\label{eqWan}
	\begin{split}
		|w^R_{m,R_i}\rangle
		&=
		\sqrt{N}
		\int
		\frac{d\beta}{2\pi i\beta}
		\beta^{-R_i}
		|\psi^R_{m,\beta}\rangle,
		\\
		\langle w^L_{m,R_i}|
		&=
		\sqrt{N}
		\int
		\frac{d\beta}{2\pi i\beta}
		\langle\psi^L_{m,\beta}|
		\beta^{R_i}.
	\end{split}
\end{align}
where \(R_i\) denotes the position of the \(i\)-th unit cell. This non-Bloch Wannier function satisfies the following relation:
\begin{align}\label{eqWanRsymmetry}
	\begin{split}
			w^R_{m,R_i}(x-R^{\prime}_i)&=\frac{1}{\sqrt{2\pi}}\int \frac{d\beta}{i\beta} \beta^{-R_i}\beta^{x-R_i^\prime}u^R_{m,\beta}(x-R_i^\prime)\\
			&=\frac{1}{\sqrt{2\pi}}\int \frac{d\beta}{i\beta} \beta^{-(R_i+R_i^\prime)}\beta^{x}u^R_{m,\beta}(x)=w^R_{m,R_i+R_i^\prime}(x).
		\end{split}
\end{align}
Similarly,
\begin{align}\label{eqWanLsymmetry}
	\begin{split}
		w^L_{m,R_i}(x-R^{\prime}_i)&=\frac{1}{\sqrt{2\pi}}\int \frac{d\beta}{i\beta} \beta^{R_i}\beta^{-(x-R_i^\prime)}u^L_{m,\beta}(x-R_i^\prime)\\
		&=\frac{1}{\sqrt{2\pi}}\int \frac{d\beta}{i\beta} \beta^{R_i+R_i^\prime}\beta^{-x}u^L_{m,\beta}(x)=w^L_{m,R_i+R_i^\prime}(x).
	\end{split}
\end{align}
Therefore, both the left and right non-Bloch Wannier functions satisfy translation symmetry. 

Moreover, the Wannier functions can be shown to satisfy a biorthogonality relation, \(\langle w^L_{m,R_i^\prime}|w^R_{n,R_i}\rangle=\delta_{mn}\delta_{R_i^\prime,R_i}\).
\begin{align}\label{eqWanbiorth}
	\begin{split}
		\langle w^L_{m,R^\prime_i}|w^R_{n,R_i}\rangle&=\int \frac{d\beta}{2\pi i\beta}\frac{d\beta^\prime}{2\pi i\beta^\prime} \langle u^L_{m,\beta^\prime}|u^R_{n,\beta}\rangle \sum_{ x}{\beta^\prime}^{R_i^\prime-x}\beta^{x-R_i}\\
		&=\int \frac{d\beta}{2\pi i\beta}\frac{d\beta^\prime}{2\pi i\beta^\prime} \langle u^L_{m,\beta^\prime}|u^R_{n,\beta}\rangle {\beta^\prime}^{R_i^\prime}\beta^{-R_i} \sum_{x}\left(\frac{\beta}{\beta^\prime}\right)^{x}\\
		&=\int \frac{d\beta}{2\pi i\beta}\langle u^L_{m,\beta}|u^R_{n,\beta}\rangle {\beta}^{R_i^\prime-R_i}=\delta_{mn}\int \frac{d\beta}{2\pi i\beta} {\beta}^{R_i^\prime-R_i}=\delta_{mn}\delta_{R_i^\prime,R_i},
	\end{split}
\end{align}
where we have used the relation shown in Eq. (\ref{eqsumseries}).

\section{The gauge-invariant part of the spread functional of the non-Bloch Wannier function}

Here we show how to decompose the spread functional of the left and right Wannier functions, \( \Omega_m \), into a gauge-invariant part, \( \Omega_m^{\mathrm{I}} \), and a gauge-dependent part, \( \Omega_m^{\mathrm{D}} \), such that \(\Omega_m=\Omega_m^I+\Omega_m^D\).
\begin{align}\label{}
	\begin{split}
		\Omega_m &= \frac{1}{N} \sum_{R_i} \Omega_{m,R_i} =\frac{1}{N}\sum_{R_i}\left[\langle w^L_{m,R_i}| \hat{x}^2| w^R_{m,R_i} \rangle- \langle w^L_{m,R_i}| \hat{x}| w^R_{m,R_i} \rangle\langle w^L_{m,R_i}| \hat{x}| w^R_{m,R_i} \rangle\right]\\
		&=\frac{1}{N}\sum_{R_i}\left[\langle w^L_{m,R_i}| \hat{x}^2| w^R_{m,R_i} \rangle- \langle w^L_{m,R_i}| \hat{x}\left(\sum_{R_i^\prime}| w^R_{m,R_i^\prime} \rangle\langle w^L_{m,R_i^\prime}|-\sum_{R_i^\prime (\neq R_i)}| w^R_{m,R_i^\prime} \rangle\langle w^L_{m,R_i^\prime}|  \right)\hat{x}| w^R_{m,R_i} \rangle\right]\\
		&=\frac{1}{N}\sum_{R_i} \langle w^L_{m,R_i}| \hat{x}\left(\mathbf{I}-\hat{W}_m\right)\hat{x}| w^R_{m,R_i} \rangle+\frac{1}{N}\sum_{R_i}\sum_{R_i^\prime(\neq R_i)}\langle w^L_{m,R_i}| \hat{x}| w^R_{m,R_i^\prime} \rangle \langle w^L_{m,R_i^\prime}| \hat{x}| w^R_{m,R_i} \rangle\\
		&=\frac{1}{N}{\rm Tr}\left\{\hat{W}_m \hat{x}\left(\mathbf{I}-\hat{W}_m\right)\hat{x}\right\}+\frac{1}{N}\sum_{R_i}\sum_{R_i^\prime(\neq R_i)}\langle w^L_{m,R_i}| \hat{x}| w^R_{m,R_i^\prime} \rangle \langle w^L_{m,R_i^\prime}| \hat{x}| w^R_{m,R_i} \rangle\\
		&=\Omega_m^I+\Omega_m^D.
	\end{split}
\end{align}

Then we prove that the projector \(\hat{W}_m = \sum_{R_i}|w^R_{m,R_i}\rangle \langle w^L_{m,R_i}|\) constructed from the non-Bloch Wannier functions is equivalent to the projector \(\hat{P}_{m} =\frac{N}{2\pi}\int \frac{d\beta}{ i\beta}|\psi^R_{m,\beta}\rangle\langle\psi^L_{m,\beta}|\) constructed from the non-Bloch eigenstates.
\begin{align}\label{eqWmeqPm}
	\begin{split}
		\hat{W}_m = \sum_{R_i}|w^R_{m,R_i}\rangle \langle w^L_{m,R_i}|=N\int \frac{d\beta}{2\pi i\beta}\frac{d\beta^\prime}{2\pi i\beta^\prime}|\psi^R_{m,\beta}\rangle \langle \psi^L_{m,\beta^\prime}|\sum_{R_i}\left(\frac{\beta}{\beta^\prime}\right)^{R_i}
	\end{split}
\end{align}
Applying Eq. (\ref{eqsumseries}), we immediately obtain
\begin{align}\label{eqWmeqPm}
	\begin{split}
		\hat{W}_m = \sum_{R_i}|w^R_{m,R_i}\rangle \langle w^L_{m,R_i}|=\frac{N}{2\pi}\int \frac{d\beta}{ i\beta}|\psi^R_{m,\beta}\rangle\langle\psi^L_{m,\beta}|=\hat{P}_{m}.
	\end{split}
\end{align}
Since \(\Omega_m^I = \frac{1}{N} \operatorname{Tr}\left\{ \hat{W}_m \hat{x} \left( \mathbf{I} - \hat{W}_m \right) \hat{x} \right\}\) and \(\hat{W}_m = \hat{P}_m\), comparing with Eq. (\ref{eqRSQGTequival}), we immediately obtain the following identity:
\begin{align}
	\Omega_m^I
	\equiv
	\mathcal{Q}^{LR}_{m,xx}
	\equiv
	\mathcal{Q}^{\rm rs}_{m,xx}.
\end{align}

\section{Wannier center and non-Blcoh Berry phase}

\begin{figure}[hbpt]
	\includegraphics[width=9.3 cm]{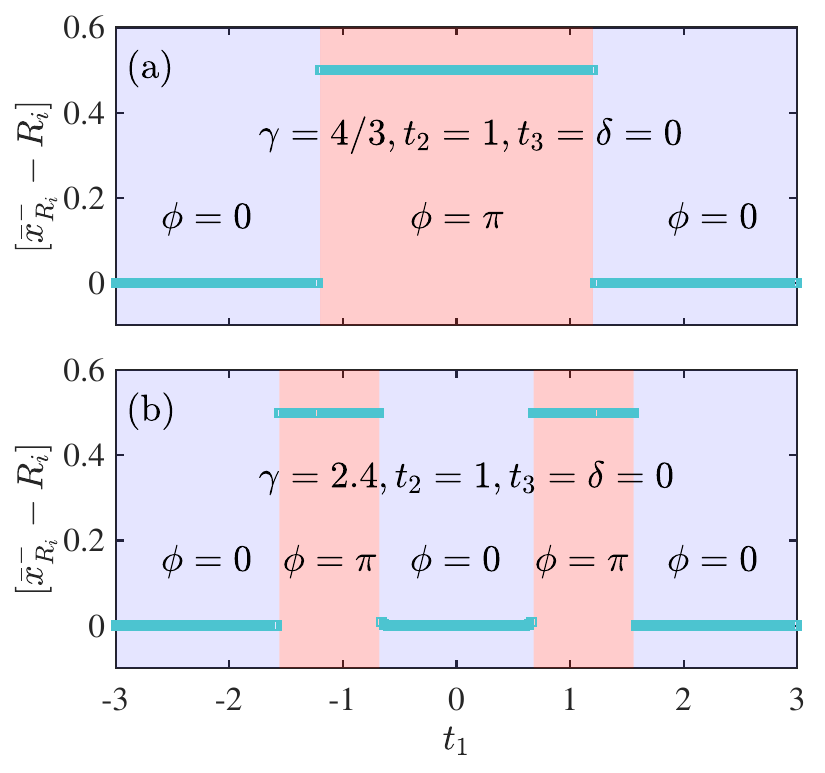}
	\caption{(a) and (b) show, for different parameter regimes, the shift of the Wannier center \(\bar{x}^{m}_{R_i}\) relative to the unit-cell position \(R_i\), i.e., \([\bar{x}_{R_i}^{m=-} - R_i]\), as a function of \(t_1\). \(m=-\) denotes the sector in which the real part of the eigenvalues is negative. The red shaded region denotes the topological phase with Berry phase \(\phi = \pi\), while the blue region corresponds to the trivial phase with \(\phi = 0\).
	}\label{figS1}
\end{figure}

Then we derive the relation between the Wannier center defined in terms of the left and right non-Bloch Wannier functions and the non-Bloch Berry connection \(\langle u^L_{m,\beta}|\partial_{k}u^R_{m,\beta}\rangle\). First, we apply the relation
\begin{align}\label{}
	\begin{split}
		x u^R_{m,\beta}\beta^{(x-R_i)} =\frac{\beta}{\partial_k\beta}\partial_k\left[u^R_{m,\beta}\beta^{(x-R_i)}\right]-\frac{\beta}{\partial_k\beta}\beta^{(x-R_i)}\partial_k u^R_{m,\beta}+R_i u^R_{m,\beta}\beta^{(x-R_i)}.
	\end{split}
\end{align}
Therefore
\begin{align}\label{eqWcenter}
	\begin{split}
		\bar{x}^m_{R_i} &= \langle w^L_{m,R_i} | \hat{x} | w^R_{m,R_i} \rangle=\int\frac{d\beta^\prime}{2\pi i\beta^\prime} \frac{d\beta}{2\pi i\beta}\sum_x u^{L*}_{m,\beta^\prime}\beta^{\prime -(x-R_i)}x u^R_{m,\beta}\beta^{(x-R_i)} \\
		&=\int\frac{d\beta^\prime}{2\pi i\beta^\prime} \frac{d\beta}{2\pi i\beta}\sum_x u^{L*}_{m,\beta^\prime}\beta^{\prime -(x-R_i)}\left\{\frac{\beta}{\partial_k\beta}\partial_k\left[u^R_{m,\beta}\beta^{(x-R_i)}\right]-\frac{\beta}{\partial_k\beta}\beta^{(x-R_i)}\partial_k u^R_{m,\beta}+R_i u^R_{m,\beta}\beta^{(x-R_i)} \right\}\\
		&=\int \frac{d\beta}{2\pi i\beta}\frac{d\beta^\prime}{2\pi i\beta^\prime} \left\{\frac{\beta}{\partial_k \beta}\partial_k \left[\langle u^L_{m,\beta^\prime}|u^R_{m,\beta}\rangle\sum_x \left(\frac{\beta}{\beta^\prime}\right)^{(x-R_i)} \right] +\left[R_i \langle u^L_{m,\beta^\prime}| u^R_{m,\beta}\rangle-\frac{\beta}{\partial_k\beta}\langle u^L_{m,\beta^\prime}|\partial_k u^R_{m,\beta}\rangle \right]\sum_x \left(\frac{\beta}{\beta^\prime}\right)^{(x-R_i)} \right\}\\
		&=\int \frac{d\beta}{2\pi i\beta} \left\{R_i-\frac{\beta}{\partial_k\beta}\langle u^L_{m,\beta}|\partial_k u^R_{m,\beta}\rangle\right\}=R_i+\frac{1}{2\pi}\int_{\rm GBZ}  i\langle u^L_{m,\beta}|\partial_{k}u^R_{m,\beta}\rangle dk,
	\end{split}
\end{align}
where \(R_i\) denotes the position of the unit cell, and \(i\langle u^L_{m,\beta}|\partial_{k}u^R_{m,\beta}\rangle\) is the Berry connection defined based on the non-Bloch Hamiltonian, and its integral over the GBZ yields the Berry phase, \(\phi = \int i\langle u^L_{m,\beta}|\partial_{k}u^R_{m,\beta}\rangle \).  Therefore, the center of the Wannier function is related to the Berry phase as
\begin{align}\label{eqWcen}
	\begin{split}
		\bar{x}^m_{R_i} &= \langle w^L_{m,R_i} | \hat{x} | w^R_{m,R_i} \rangle=R_i+\frac{\phi}{2\pi}.
	\end{split}
\end{align}
In Figure~\ref{figS1}, we present the shift of the Wannier center \(\bar{x}^m_{R_i}\) relative to the unit-cell position \(R_i\) under different parameter regimes, which is consistent with the prediction of Eq.~(\ref{eqWcen}).

\section{More results}
\begin{figure}[hbpt]
	\includegraphics[width=11.8 cm]{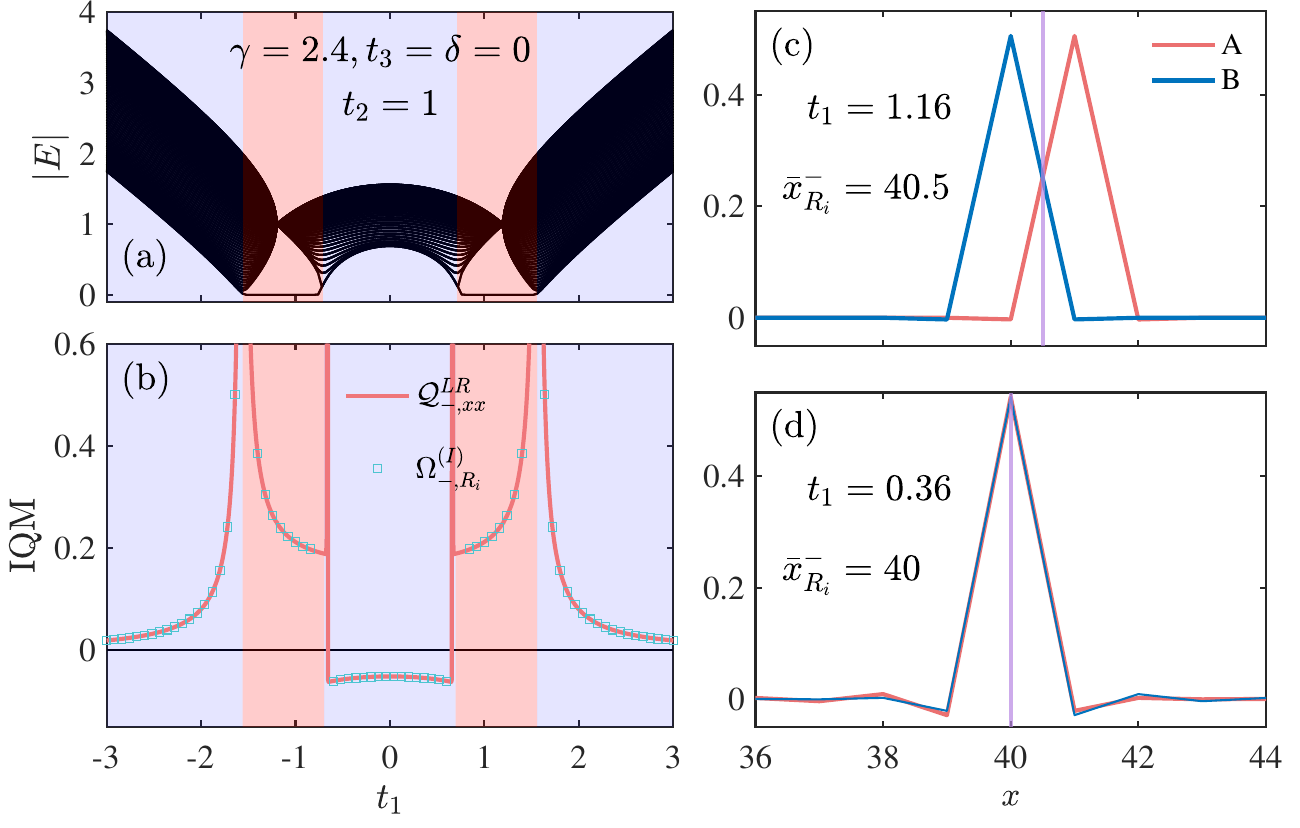}
	\caption{(a) and (b) show, respectively, the absolute value of the energy spectrum and the IQM as functions of \( t_1 \) for the same set of parameters. The red (blue) shaded regions indicate the topological (trivial) phase. (c) displays the distribution of  \( f(x) = [w^{L(I)}_{-,R_i}(x)]^* w^{R(I)}_{-,R_i}(x) \) at \( t_1 = 1.16 \), corresponding to the topological phase. (d) is the same as (c), but for \( t_1 = 0.36 \), corresponding to the trivial phase. The vertical lines in (c) and (d) mark the Wannier centers \( \bar{x}^{-}_{R_i} = \langle w^{L(I)}_{-,R_i} | \hat{x} | w^{R(I)}_{-,R_i} \rangle \), defined using the left and right Wannier functions.
	}\label{figS4}
\end{figure}
In Figure~\ref{figS4}, we present the results for another set of parameters, showing the behavior of the non-Hermitian SSH model as \(t_1\) varies. In this case, two topological regions (highlighted by red shading) emerge. The results of the IQM \(Q^{LR}_{-,xx}\) are consistent with the Wannier spread functional \(\Omega^{(I)}_{-,R_i}\). Moreover, in the trivial phase around \(t_1 = 0\), the value of the Wannier spread functional \(\Omega^{(I)}_{-,R_i}\) becomes negative. Fig \ref{figS4}(c) and (d) display the spatial distribution of the product of the left and right Wannier functions in the topological and trivial phases, respectively. From the distribution shown in Fig.~\ref{figS4}(d), it is clear that the profile of \(f(x)= [w^{L(I)}_{-,R_i}(x)]^* w^{R(I)}_{-,R_i}(x)\) leads to a negative  \( \Omega_{-,R_i}^{(I)} = \sum_{x} f(x) (x - \bar{x}^-_{R_i})^2\).

\begin{figure}[hbpt]
	\includegraphics[width=12.8 cm]{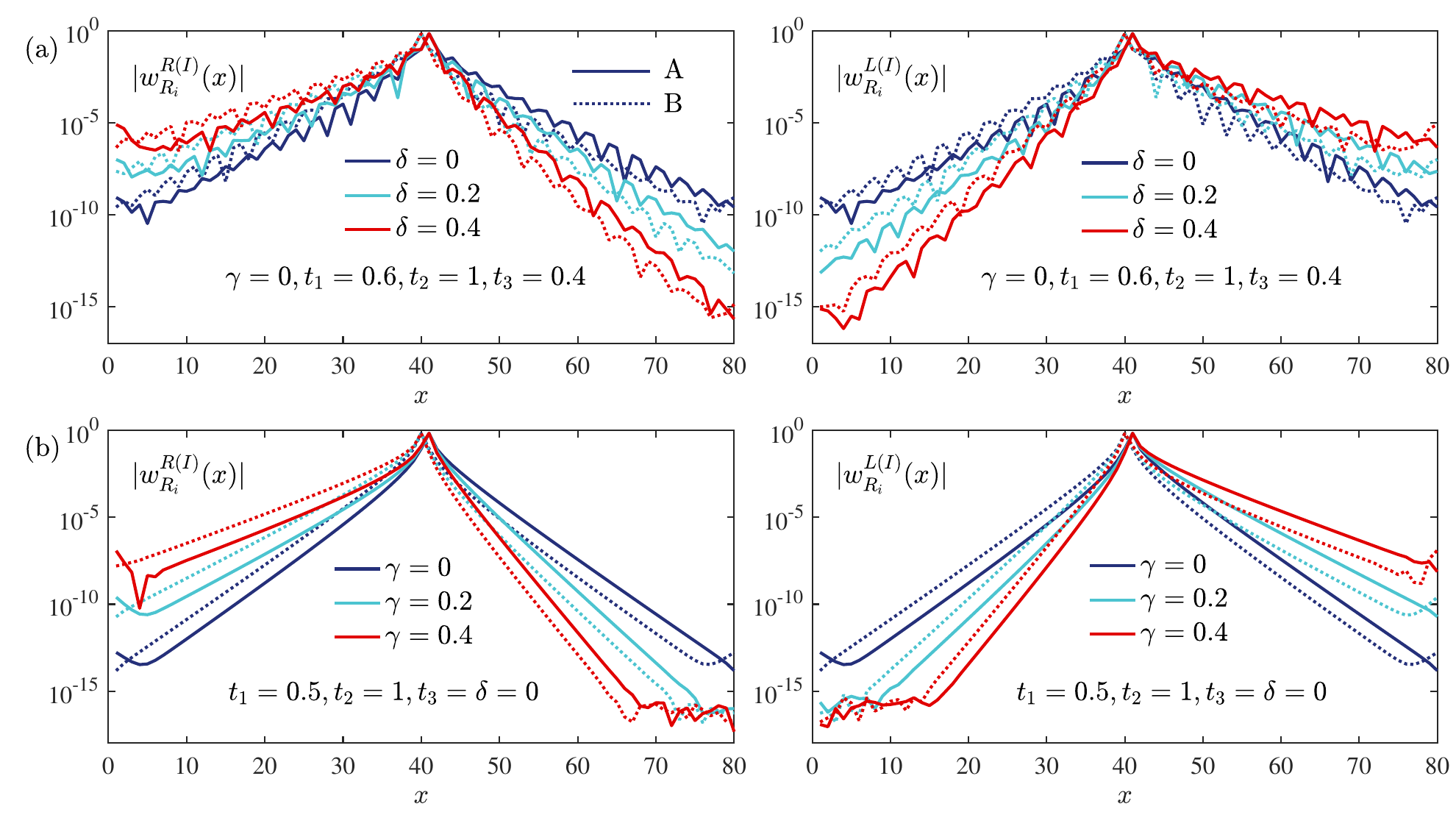}
	\caption{(a) Spatial profiles of \(|w^{R(I)}_{m,R_i}\rangle\) (left panel) and \(\langle w^{R(I)}_{m,R_i}|\) (right panel) for different values of \( \delta \), shown on a logarithmic scale along the \( y \)-axis. (b) Spatial profiles of \(|w^{R(I)}_{m,R_i}\rangle\) (left panel) and \(\langle w^{R(I)}_{m,R_i}|\) (right panel) for different values of \( \gamma \) in another parameter regime. In both (a) and (b), solid and dashed lines denote the distributions on the two sublattices, respectively. The cases \( \delta = 0 \) in (a) and \( \gamma = 0 \) in (b) correspond to the Hermitian limit.
	}\label{figS5}
\end{figure}

In Figure~\ref{figS5}, we present the spatial distributions of the left and right projected-position Wannier functions under different parameter regimes, as the non-Hermitian parameter varies. We can clearly see that the behavior of the left Wannier function as the non-Hermitian parameter varies is opposite to that of the right Wannier function.

